\documentclass[preprint]{elsarticle}
\usepackage{epsf,colordvi,color,graphicx,epsfig}
\usepackage{amsfonts,amssymb,amsmath,wasysym,color}

\begin{document}

\title{Coexisting patterns of population oscillations: the degenerate Neimark
Sacker bifurcation as a generic mechanism}

\author[da]{Christian Guill}\corref{cor1}
\ead{guill@fkp.tu-darmstadt.de}

\author[da]{Benjamin Reichardt}
\ead{ben.reich@web.de}

\author[da]{Barbara Drossel}
\ead{drossel@fkp.tu-darmstadt.de}

\author[uk]{Wolfram Just}
\ead{w.just@qmul.ac.uk}

\cortext[cor1]{Corresponding author}

\address[da]{Institut f\"ur Festk\"orperphysik, TU Darmstadt, 
Hochschulstrasse 6, 64289 Darmstadt, Germany}

\address[uk]{School of Mathematical Sciences, Queen Mary University
of London, Mile End Road, London E14NS, UK}

\begin{abstract}
We investigate a population dynamics model that exhibits a Neimark
Sacker bifurcation with a period that is naturally close to 4. Beyond
the bifurcation, the period becomes soon locked at 4 due to a strong
resonance, and a second attractor of period 2 emerges, which coexists
with the first attractor over a considerable parameter range.  A
linear stability analysis and a numerical investigation of the second
attractor reveal that the bifurcations producing the second attractor
occur naturally in this type of system.
\end{abstract}


\maketitle

\section{Introduction}

The population dynamics of biological species that produce offspring
once in their lifetime (semelparous species) provides a generic system 
that can display a Neimark Sacker bifurcation. A remarkable example 
is the pacific sockeye salmon (\textit{Oncorhynchus nerka}), which 
returns from the ocean to spawn in its native
lake and dies afterwards. The sockeye fry spend one season in the
lake, feeding on zooplankton and being eaten by predator fish such as
rainbow trout, and then migrate to the ocean, from where they return
to spawn at the age of 3, 4, or 5 years. Several sockeye populations display large-amplitude oscillations with a period corresponding to the dominant generation time of these fish, a phenomenon generally referred to as cyclic dominance \cite{ricker}. In contrast 
to conventional predator-prey systems, the dynamics of the sockeye-trout 
system in the lake is piecewise continuous in time, due to the salmon 
migrating to the ocean and returning at the age of 4 ($\pm 1$) years. 
This mechanism, which puts the model in the context of delay
dynamical systems, will turn out to be one of the crucial ingredients.
The second one is the external periodic drive caused by the yearly rhythm
of seasons. The ratio of these two time scales is thus fixed by
an internal mechanism which turns out to be vital for the observed
phenomena.

While conventional predator-prey systems typically display a Hopf 
bifurcation \cite{rosenzweig}, our piecewise continuous system displays a Neimark Sacker
bifurcation \cite{unserlachspaper}, i.e., an oscillatory instability in
externally driven systems. Since the dominant life span of the 
salmon is 4 years, the period of the oscillation is naturally close to 4, 
and when the control parameter is further increased beyond the bifurcation, 
the period becomes locked exactly at 4 due to the strong resonance 
familiar from the theory of the Neimark Sacker bifurcation \cite{kuznetsov}. 

Resonances and the associated synchronisation phenomena are common
features in driven systems, with widespread applications in science
and technology (cf. \cite{sync} for a recent review).
But unlike strong resonances one normally
requires a fine tuning of parameters, the subtle
synchronisation phenomena like phase locking or universal scaling
behaviour commonly related with the quasiperiodic transition to
chaos \cite{qpren,dsdim} may be sensitive to perturbations.
In contrast, strong resonances in Neimark Sacker bifurcations
generate very stable periodic motion over large regions in parameter
space, and thus can provide a vital mechanism for 
cyclic dominance in biological systems where
noise and parameter fluctuations are prevalent.

In this paper, we will explore further a model for the population
dynamics of the pacific sockeye salmon. We will
show that it can display a second attractor, which has the period 2
and coexists with the attractor of period 4. Due to the strong
resonance, the second attractor is clearly visible in bifurcation
diagrams where only one of the 6 dimensions of the system is plotted
as function of a control parameter. By using computer simulations, we
will show that the second attractor arises through a flip bifurcation
followed by a subcritical Neimark Sacker bifurcation, and by using a
theory that is linearised around the fixed point as well as around the
attractor of period 4, we will show that the flip bifurcation can 
naturally occur soon after the first Neimark Sacker bifurcation.

\section{The model}

The dynamics of the biomass of sockeye fry $s_n(t)$, of their
predator $p_n(t)$, and of their zooplankton food $z_n(t)$ 
in year number $n$ during the growth season from spring 
($t=0$) to fall ($t=T$) are given by the following 
set of coupled differential equations \cite{unserlachspaper}:
\begin{eqnarray} \label{dgl}
\frac{d}{dt} s_n (t) &=& \lambda g_{sz}(s_n(t), z_n(t))- g_{ps}(p_n(t),s_n(t)) 
- d_s\cdot s_n(t) \nonumber\\
\frac{d}{dt} z_n (t) &=& z_n(t)\cdot\left(1-\frac{z_n(t)}{K_n}\right)
-g_{sz}(s_n(t), z_n(t)) \\
\frac{d}{dt} p_n (t) &=& \lambda g_{ps}(p_n(t),s_n(t)) - d_p\cdot p_n(t) 
\nonumber
\end{eqnarray}
The terms in these equations represent the biomass changes due to eating 
and being eaten, and the losses due to respiration and death. 
The two feeding terms are given by
\begin{eqnarray}
g_{sz}(s, z) &=& 
a_{sz}\frac{z\cdot s}{1+c_s\cdot s+z}  \nonumber \\
g_{ps}(p,s) &=&
a_{ps}\frac{s\cdot p}{1+c_p\cdot p+s} 
\end{eqnarray}
These functional forms include saturation at high prey densities, and a
predator interference term in the denominator (Beddington functional
response \cite{beddington,skalski}). These equations of motion determining the
time evolution of the biomass during a single season are supplemented
with matching conditions used to determine the biomasses of the species at 
the beginning of the next season from their values at the end of the previous 
season(s). To keep the model as simple as possible we assume that
no major change in the population of predators occurs.
The zooplankton level at the end of one year has no effect on the following year \cite{zoo1}, but its saturation value in the next season, $K_{n+1}$, shows a
dependence on the nutrient input due to the number of
spawning adults, and thereby on $ s_{n+1}(0)$ \cite{zoo2,zoo3}. 
\begin{eqnarray}
z_{n+1}(0)&=& K_{n+1} = K_0+
\left(\delta\frac{s_{n+1}(0)}{\delta_0+s_{n+1}(0)}\right) \label{zmatch}\\
p_{n+1}(0)&=&p_n(T) \label{pmatch}
\end{eqnarray}
Most importantly the matching condition for the sockeye fry population
contains the proportion $\epsilon_1$ and $\epsilon_2$ 
of surviving sockeye that return to their native 
lakes at the age of 3 and 5 to spawn and die:
\begin{equation}\label{smatch}
s_{n+1}(0)=\gamma[(1-\epsilon_1-\epsilon_2) s_{n-3}(T)+\epsilon_1 s_{n-2}(T)
+\epsilon_2 s_{n-4}(T)]
\end{equation}
On the one hand the model may be considered as a very simple example of a 
piecewise smooth
dynamical system \cite{supermario}, where the discontinuities are contained 
in the matching condition between seasons. On the other hand the model
may be considered as a periodically driven non autonomous system where
the external drive is provided by the seasons. It is therefore an
obvious approach to consider the stroboscopic map. If we denote by
$\Phi_{[0,T]}$ the flow of the differential equation (\ref{dgl}) the
biomasses at the beginning and at the end of the year $n+1$
are related by
\begin{equation} \label{strobmap}
(s_{n+1}(T),p_{n+1}(T),z_{n+1}(T))=\Phi_{[0,T]}(s_{n+1}(0),p_{n+1}(0),z_{n+1}(0))
\end{equation}
The condition (\ref{zmatch}) tells us that the zooplankton is entirely
determined by sockeye population so that $z_n$ drops from Eq.(\ref{strobmap})
as a dynamical variable. Taking Eqs.(\ref{pmatch}) and (\ref{smatch}) into
account the remaining biomass evolution can be written as
\begin{eqnarray}\label{map}
s_{n+1}(T)&=&h_s(\epsilon_1 s_{n-2}(T)+
(1-\epsilon_1-\epsilon_2) s_{n-3}(T)
+\epsilon_2 s_{n-4}(T),p_n(T))\nonumber \\
p_{n+1}(T)&=&h_p(\epsilon_1 s_{n-2}(T)+
(1-\epsilon_1-\epsilon_2) s_{n-3}(T)
+\epsilon_2 s_{n-4}(T),p_n(T)) \, .
\end{eqnarray}
Thus we arrive at a six dimensional time discrete dynamical system.
The computation of the right hand sides, $h_s$ and $h_p$,
requires the integration of the
equations of motion, Eq.(\ref{dgl}), a task which can be accomplished 
easily by numerical means. However, the structure of the map (\ref{map}),
induced by the matching condition (\ref{smatch}) will turn
out to be crucial for the properties of the population dynamics.

\section{The main bifurcation generating cyclic dominance}

It is our main purpose to pinpoint an underlying generic
mechanism which creates stable cyclic motion among the populations
in our model. For that purpose we employ a brief bifurcation
analysis for realistic combinations of the parameters of the
model. Given the fairly large number of parameters in our model
and the fairly large number of degrees of freedom we just resort
to plain numerical simulations. In fact, employing for our
specific purpose numerical
continuation tools like matcont \cite{matcont}
our auto \cite{auto} would be using a sledge-hammer to crack a nut.
In addition, our specific numerical study mimics approaches which
could be used in evaluating empirical data, and thus could be a
guide for their analysis.

\begin{figure}[h!t!b!p]
 \begin{center}
  \includegraphics[angle=0,scale=1]{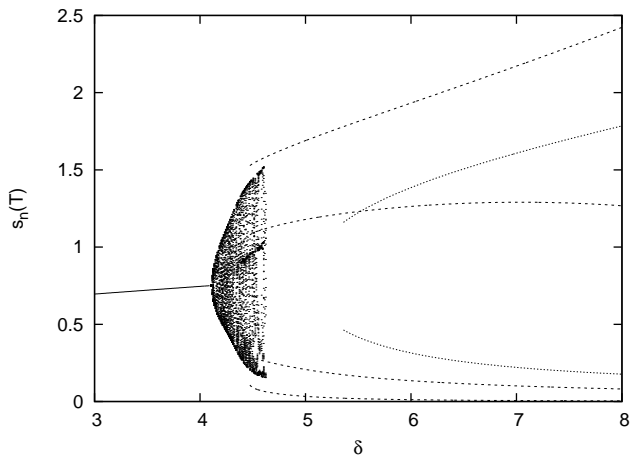}
 \end{center}
\caption{Bifurcation diagram of the biomass of the sockeye fry at the 
end of their first year. Parameter values used for this diagram are: 
$\lambda=0.85$, $a_{sz}=10$, $a_{ps}=0.2$, $c_s=1$, $c_p=0.2$, $d_s=1.1$, 
$d_p=0.3$, $\gamma=0.2$, $\epsilon_1=0$, $\epsilon_2=0.1$, $K_0=2.5$, 
and $\delta_0=2$.}
\label{fig:bifurcation}
\end{figure}

Figure \ref{fig:bifurcation} shows the bifurcation diagram for a 
realistic set of parameter values. For each value of the parameter 
$\delta$, several initial conditions were drawn at random from uniform probability distributions on a 6-dimensional cube $(0;10]$. When coexisting attractors were found, these were numerically continued until they became unstable. After a transient 
time of 5000 years (longer close to bifurcations), the values for 
$s_n(T)$ were plotted for 20 additional years. As $\delta$ increases, 
the fixed point becomes unstable due to a Neimark Sacker bifurcation, 
and a quasiperiodic attractor arises. At $\delta\simeq 4.47$, trajectories 
can become locked exactly at the period 4, but the quasiperiodic 
attractor remains stable until $\delta\simeq 4.6$. This picture is
consistent with a Neimark Sacker bifurcation close to a strong
resonance of order 1:4, which is known to display quite a subtle
bifurcation structure \cite{kuznetsov}, including the creation of
homoclinic tangles.

At $\delta \simeq 5.36$, a second attractor with period 2 appears. 
Such a feature is normally not related with the generic strong resonant
Neimark Sacker bifurcation mentioned above. The
two arms of this attractor appear as if they are created
as an unstable period two orbit at $\delta \simeq
5.1$ via a flip bifurcation from the already unstable fixed point. 
The unstable arms then become stable at
$\delta \simeq 5.36$ when the complex conjugate pair of eigenvalues
moved back into the unit circle. Even without continuation tools
tracking saddle orbits we are able to vindicate this picture by
evaluating the size of the basin of attraction of the period two attractor
and by monitoring the transient dynamics.

\begin{figure}[h!t!b!p]
 \begin{center}
  \includegraphics[angle=0,scale=1]{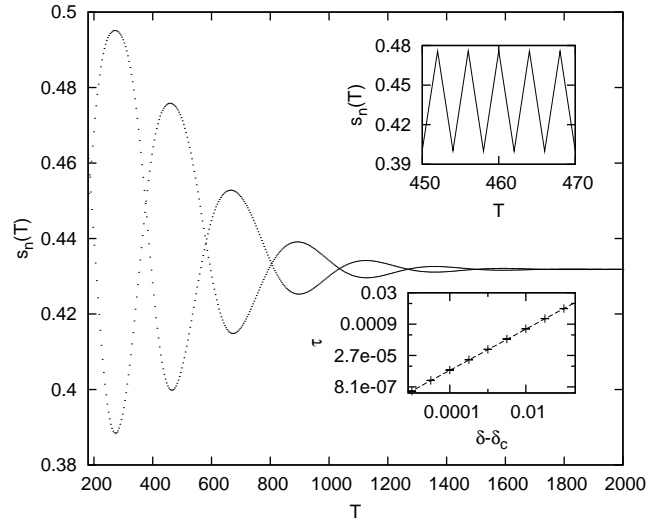}
 \end{center}
\caption{Trajectory approaching the period 2 attractor. $\delta=5.45$, all other parameters are as in figure 1. Only every second data point is plotted. Upper inset: close-up of the trajectory. Lower inset: decay rate of the oscillations when approaching the instability. The dashed line is a linear function fitted to the numerically determined decay rates, yielding $\delta_c=5.356111\pm 5\cdot 10^{-7}$.}
\label{fig:trajectorie}
\end{figure}

\begin{figure}[h!t!b!p]
 \begin{center}
  \includegraphics[angle=0,scale=1]{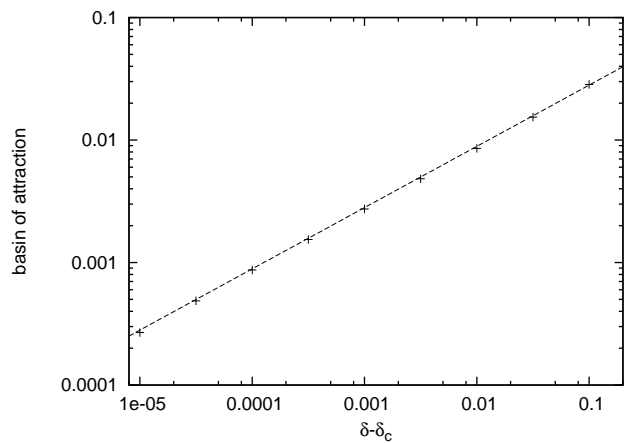}
 \end{center}
\caption{Size of the basin of attraction of the period 2 attractor. 
Parameters are as in figure \ref{fig:bifurcation}, and 
$\delta_c=5.356111$. The dashed line is a power law with exponent $1/2$.}
\label{fig:basin}
\end{figure}

Figure \ref{fig:trajectorie} shows an example of the transient dynamics 
in the stable period two regime. Since only every other iteration is
displayed the transient consists of a damped oscillation with a period 
close to 4. By measuring the decay rate $\tau$ of the oscillations around the stable
period two orbit for different values of $\delta$ we determined the bifurcation value $\delta_c$ with an accuracy of $10^{-6}$. Sufficiently close to the
bifurcation, $\tau$ goes linearly to 0 as the instability is approached (lower
inset in Figure \ref{fig:trajectorie}). We then estimated the diameter of the
basin of attraction by using a
one dimensional cross section along the direction of the first
salmon population. Initial conditions were distributed along this
cross section, and those trajectories which return to the
period two state determine the extent of the basin along this
direction in phase space. Figure \ref{fig:basin} shows the result
for the size of the basin, and the characteristic square root law 
with regard to the distance from the bifurcation point is recorded.
Thus the bifurcation which causes the stability of the period two
state is clearly a subcritical bifurcation. 

All these results demonstrate that
the second attractor emerges due to a subcritical Neimark Sacker
bifurcation, and this secondary Neimark Sacker bifurcation is
intimately related to the bifurcation creating the stable period
four attractor. We are now going to uncover the underlying mechanism
by analytical considerations.

\section{Linear stability analysis}

A linear stability analysis of the dynamical system close to the first
Neimark Sacker bifurcation reveals that the scenario outlined in the
previous section is indeed a natural scenario for this type of system.
The continuous population dynamics during the season, together with
the matching conditions applied between two seasons, can be viewed as
a Poincar\'e map, Eq.(\ref{map}), giving the sockeye and predator 
biomasses at the end of one year as function of the biomasses at the 
end of the previous years.  Close to the fixed point $(s^*,p^*)$, 
the dynamics can be approximated by linear terms. Denoting the distance 
of the biomasses from their fixed point
value by $\delta s_n=s_n(T)-s^*$ and  $\delta p_n=p_n(T)-p^*$, 
the linear approximation of Eq.(\ref{map}) has the form
\begin{equation}\label{linmap}
\left(\begin{array}{c}\delta s_{n+1} \\ \delta s_{n} \\\delta s_{n-1}
\\\delta s_{n-2}\\\delta s_{n-3}\\\delta p_{n+1}
\end{array}\right)
=
\left(\begin{array}{cccccc}
0&0&\alpha \epsilon_1&\alpha(1-\epsilon_1-\epsilon_2) &\alpha
\epsilon_2 &\beta\\
1&0&0&0&0&0\\
0&1&0&0&0&0\\
0&0&1&0&0&0\\
0&0&0&1&0&0\\
0&0&\bar{\alpha}\epsilon_1&\bar{\alpha}(1-\epsilon_1-\epsilon_2)&
\bar{\alpha}\epsilon_2 &\bar{\beta}
\end{array}\right)
\left(\begin{array}{c}\delta s_{n} \\ \delta s_{n-1} \\\delta s_{n-2}
\\\delta s_{n-3}\\\delta s_{n-4}\\\delta p_{n}
\end{array}\right)
\end{equation}
where
\begin{eqnarray}\label{param}
\alpha= \frac{\partial h_s(s^*,p^*)}{\partial s^*}, &\quad&
\beta=\frac{\partial h_s(s^*,p^*)}{\partial p^*} \nonumber \\
\bar{\alpha}= \frac{\partial h_p(s^*,p^*)}{\partial s^*}, &\quad&
\bar{\beta}=\frac{\partial h_p(s^*,p^*)}{\partial p^*} 
\end{eqnarray}
denote the partial derivatives of the Poincaré map at the fixed point.
These parameter values can in principle be obtained by a numerical
evaluation of the population dynamics, Eq.(\ref{dgl}), close to the 
fixed point. Any consistent underlying dynamical model should give
a negative value of the parameter $\beta$, since an increase in trout
biomass leads to a decrease in salmon biomass. The other parameters
in Eq.(\ref{param}) have to be non-negative, since more parents imply 
more offspring and since an increase in salmon biomass leads to an 
increase in trout biomass. However, it is notoriously difficult to
derive such conditions from equations of motion like Eq.(\ref{dgl}).

The eigenvalues of the matrix in Eq.(\ref{linmap})
determine the nature of the dynamics
near the bifurcation. When all eigenvalues have an absolute value
smaller than 1, the fixed point is stable, and the dynamics converges
to this fixed point. When the absolute value of one or more
eigenvalues is larger than 1, the fixed point is unstable, and the
dynamics approaches a different attractor. 
The eigenvalues $\mu$ satisfy the characteristic equation
\begin{equation}\label{chargl}
\mu^5(\bar{\beta}-\mu)+[\epsilon_1 \mu^2 +(1-\epsilon_1-\epsilon_2) \mu
+ \epsilon_2][\beta \bar{\alpha}-\alpha(\bar{\beta}-\mu)]=0 \, .
\end{equation}
In order to understand the possible bifurcation scenarios, we
first consider the case that only the parameters $\alpha$, $\beta$ and 
$\bar{\beta}$ are nonzero. This means that the four salmon lines are 
independent from each other, $\epsilon_1=\epsilon_2=0$, and from the 
trout, $\bar{\alpha}=0$. The eigenvalues of the matrix 
are $\bar{\beta}$, 0, and the four fourth roots of $\alpha$. 
The eigenvalue $\bar{\beta}$ characterises the time evolution of the trout, 
which is decoupled from the salmon in the considered case. Global stability
criteria require this eigenvalue to be smaller than one, $0\leq \bar{\beta}<1$.
The eigenvalue $0$ is due to the fifth column of the matrix being redundant 
if $\epsilon_2=0$. The four fourth roots of $\alpha$ describe the salmon 
dynamics in the vicinity of the fixed point. 
The sequence $\delta s_n$ has trivially the period 4 and
simply iterates the initial four values, with an amplitude decreasing
for $\alpha<1$ and increasing otherwise.  When $\alpha$ is increased 
from a value smaller than 1 to a value larger than 1, all four eigenvalues
$\alpha^{1/4}$ cross the unit circle simultaneously, and the fixed point
becomes unstable.  This degeneracy is lifted when the other parameters,
$\epsilon_1$, $\epsilon_2$, and $\bar{\alpha}$,
are made nonzero. As long as these
parameters are not large, one can expect the four main eigenvalues to
remain close to the unit circle, implying a
(possibly damped) oscillation with a period close to 4. 
If the complex conjugate pair of eigenvalues is the first one to cross 
the unit circle, a Neimark Sacker bifurcation occurs. If the negative 
eigenvalues crosses the unit circle first, a flip bifurcation occurs. 
For the parameter values chosen in the simulation 
(Figure \ref{fig:bifurcation}), the Neimark Sacker bifurcation is 
observed first. 

The entire bifurcation scenario can be computed from 
Eq.(\ref{chargl}) using the Schur-Cohn-Jury criterion \cite{Jury},
but the main mechanism which is at the heart of the bifurcation discussed
in the previous section can be already uncovered if we resort to a
perturbative approach. In order to gain some analytical insights, 
we assume that the parameters $\epsilon_1$, $\epsilon_2$, and
$\bar{\alpha}$ are so small that the change in $\mu$ can be calculated 
approximately by performing a first order
Taylor expansion in $\alpha-1$, close to the bifurcation. 
By implicit differentiation of Eq.(\ref{chargl}) we obtain for
the relative change of the eigenvalues close to the unit circle
\begin{equation}\label{evex}
\frac{\delta \mu}{\mu_0} = \frac{\alpha-1}{4}-
\left\{
\begin{array}{lcl}
\displaystyle
\frac{(-\beta)\bar{\alpha}}{4(1-\bar{\beta})} & \mbox{ if }& \mu_0=1 \\
\displaystyle
\frac{\epsilon_1+\epsilon_2}{2}-\frac{(-\beta)\bar{\alpha}}{4(1+\bar{\beta})} 
& \mbox{ if }& \mu_0=-1  \\
\displaystyle
\frac{\epsilon_1+\epsilon_2}{4}-\frac{(-\beta)\bar{\alpha}\bar{\beta}}
{4(1+\bar{\beta}^2)}\mp \frac{i}{4}\left(\epsilon_1-\epsilon_2 +
\frac{(-\beta)\bar{\alpha}}{4(1+\bar{\beta}^2)}\right)  
& \mbox{ if }& \mu_0=\pm i
\end{array} \right.
\end{equation}
where $\mu_0$ denotes the eigenvalue of the unperturbed case.
The real part of this ratio $\delta\mu/\mu_0$ determines the stability 
properties of the eigenmode. As long as the parameter $\bar{\beta}$ is 
not too small, i.e. as long as the decay of the trout population is
not massive, the negative contribution to the positive real eigenvalue
(the case $\mu_0=1$) dominates,
and this mode remains stable on changing the
bifurcation parameter $\alpha-1$. Thus no saddle node bifurcation is
expected to occur. The order of the other two cases, a bifurcation caused
by a negative eigenvalue (the case $\mu_0=-1$) and a 
strongly resonant Neimark Sacker bifurcation (the case $\mu_0=\pm i$)
is now determined by the balance between the unfolding parameters.
If $\epsilon_1+\epsilon_2$ is sufficiently large compared to $\bar{\alpha}$,
the Neimark Sacker bifurcation appears first, as the negative contribution 
to the second case in Eq.(\ref{evex}) is larger (in modulus) than the
negative contribution to the third case. In quantitative terms Eq.(\ref{evex})
yields for the suppression of the saddle node bifurcation the inequality
\begin{equation}\label{snsup}
\frac{(-\beta)\bar{\alpha}}{1-\bar{\beta}^2} > \epsilon_1+\epsilon_2
\end{equation}
whereas the Neimark Sacker bifurcation precedes the flip bifurcation
if
\begin{equation}\label{flns}
\epsilon_1+\epsilon_2 > \frac{(-\beta)\bar{\alpha} (1-\bar{\beta})}
{(1+\bar{\beta})(1+\bar{\beta}^2)}
\end{equation}
is satisfied (remember that $\beta<0$ and $0<\bar\beta<1$).

These considerations show that as long as the coupling between the
different salmon lines and between the salmon and the trout is small
compared to the coupling between salmon parents and offspring
(i.e. $\epsilon_1\ll 1$, $\epsilon_2\ll 1$, and $\bar{\alpha}\ll 1$), the
oscillation period of the population is close to four.
If the coupling between the salmon lines is finite and beyond a threshold value determined by
the small coupling between the salmon and the trout (i.e. by the
parameter $\bar{\alpha}$, cf. Eq.(\ref{flns}))
the period four oscillating mode becomes unstable first
as the bifurcation parameter $\alpha$ is increased,
and the complex conjugate eigenvalues close to the imaginary axis
are pushed towards the unit circle. The complex conjugate 
eigenvalues are therefore the first ones to cross the unit circle 
followed by the eigenvalue on the negative real axis. But we require as well
a decay rate for trouts which is not excessively large, Eq.(\ref{snsup}),
to prevent the saddle node bifurcation to occur.
In fact, in the limit $\bar{\beta}\rightarrow 0$ the two conditions,
Eqs.(\ref{snsup}) and (\ref{flns}) become mutually exclusive. Thus
the dynamics of the predators play an important role and the
scenario would not appear if an adiabatic elimination of the
trout population would have been an option. The scenario found in
the computer simulations, where first a Neimark Sacker bifurcation
occurs and then a flip bifurcation, is thus a generic scenario in this
type of system. Clearly, our linear analysis cannot keep track of the 
stability properties of the period two orbit created
in the secondary flip bifurcation. Certainly this orbit inherits
its stability properties from the bifurcating fixed point, but
we cannot predict without a tedious normal form calculation
whether the two unstable complex conjugated
eigenvalues move back into the unit circle, although such a
re-entrance phenomenon is not uncommon. While we observed the
subcritical Neimark Sacker bifurcation on the period-2 orbit for a
broad set of parameter values, it did not occur in other regions of
parameter space. But the presented analytical considerations
have shown that the stable period two and period four oscillations
are caused by the same degenerate Neimark Sacker bifurcation, and
we were able as well to derive criteria for the occurrence of this
scenario.

 \section{Conclusions}

We have shown that a three-species population dynamics model can
serve as a model system for observing the coexistence of two
attractors, one of period four and one of period two. This is remarkable,
since this type of scenario is of codimension 3, requiring a
Neimark-Sacker bifurcation occurring with a period close to four, and
a third eigenvalue crossing the unit cycle for a nearby value of the
control parameter. We have identified conditions such that
the bifurcation of high codimension does not require fine tuning
of additional parameters. First of all, the external stimulus
of yearly season and the return of species for spawning provides
the strong resonance condition of the Neimark Sacker bifurcation and
the reduction to a time discrete dynamical model. The second important
ingredient is a predator population which actively takes part
in the dynamics and cannot be taken into account on a plain adiabatic
description. Finally a weak coupling between the different spawning
populations is required. While the zooplankton provides the source
for keeping the nonequilibrium dynamics going, it does not independently
take part in the dynamics but depends solely on the current salmon population. While our considerations have
been based on a simple but realistic model, Eq.(\ref{dgl}), we should
emphasise that the conclusions have a fairly broad remit as the
time discrete dynamics (\ref{map}) only takes a few generic features
of the model into account. Population models that have discrete generations
can thus represent a natural situation for a scenario that was
previously expected to require very improbable special conditions.

At a more fundamental abstract level the model proposed in this
contribution is at the interface between piecewise smooth dynamical systems
and driven delay dynamics, both being fields of active research in
applied dynamical systems theory \cite{supermario,dcs}. 
Although being a standard aspect of bifurcation 
theory, the mechanism presented here could have a wider impact for
the study of robust periodic motion and synchronisation in a large class
of experimentally relevant setups in science and engineering.

\end{document}